\documentclass[10pt]{article}
\usepackage{spconf,amsmath,graphicx}
\usepackage{mathtools, amsfonts, url, cite, hyperref, subcaption, booktabs}
\usepackage{subfloat}

\title{Highly controllable diffusion-based any-to-any voice conversion model with frame-level prosody feature}
\name{Kyungguen Byun, Sunkuk Moon, and Erik Visser}
\address{Qualcomm Technologies, Inc.}
\begin{document}
%
\maketitle
\begin{abstract}
We propose a highly controllable voice manipulation system that can perform any-to-any voice conversion (VC) and prosody modulation simultaneously. State-of-the-art VC systems can transfer sentence-level characteristics such as speaker, emotion, and speaking style. However, manipulating the frame-level prosody, such as pitch, energy and speaking rate, still remains challenging. Our proposed model utilizes a frame-level prosody feature to effectively transfer such properties. Specifically, pitch and energy trajectories are integrated in a prosody conditioning module and then fed alongside speaker and contents embeddings to a diffusion-based decoder generating a converted speech mel-spectrogram. To adjust the speaking rate, our system includes a self-supervised model based post-processing step which allows improved controllability. The proposed model showed comparable speech quality and improved intelligibility compared to a SOTA approach. It can cover a varying range of fundamental frequency (F0), energy and speed modulation while maintaining converted speech quality. 

\end{abstract}


\begin{subfigures}
\begin{figure*}
\includegraphics[width=0.53\textwidth]{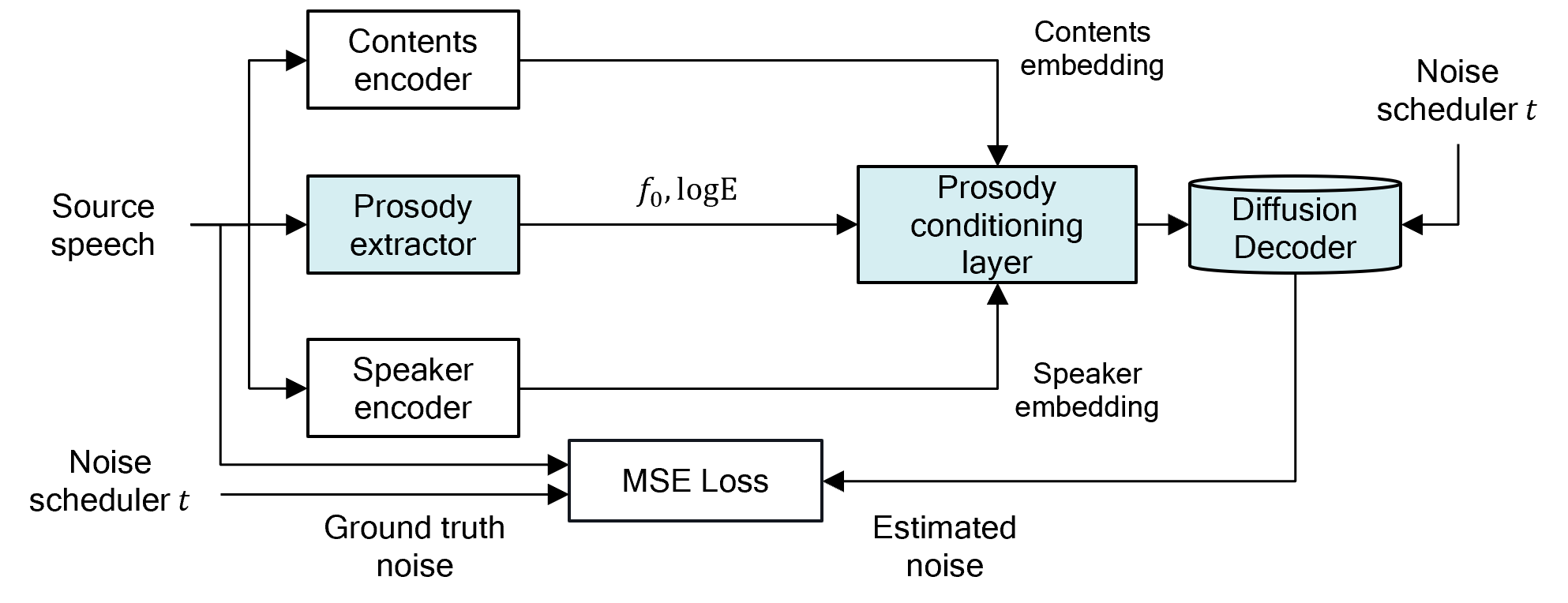}
\centering
\caption{\label{first} Training process of the proposed model.} 
\end{figure*}
\begin{figure*}
\includegraphics[width=0.75\textwidth]{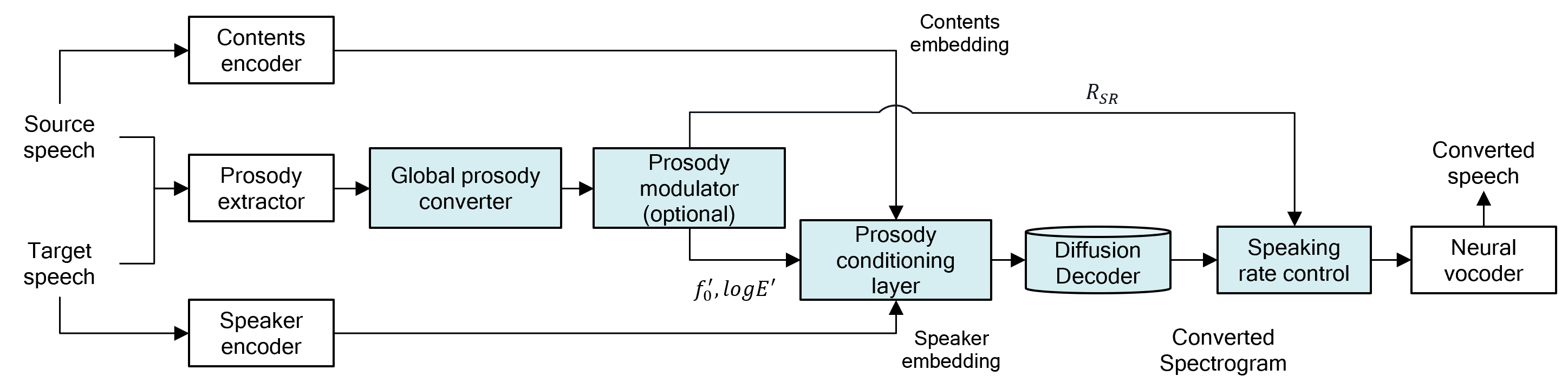}
\centering
\caption{\label{second} Inference process of the proposed model. Proposed modules are highlighted with colors.} 
\end{figure*}
\end{subfigures}
\vspace{-10pt}
\section{Introduction}
\label{sec:intro}
\label{sec:pagestyle}

The goal of voice conversion (VC) is to generate a target-like voice while preserving the linguistic content of the source speech. This technique has the potential to be used in a wide range of applications, such as XR (eXtended Reality), VR (Virtual Reality), and data augmentation for low-resource domains. Recently, the converted speech quality of voice conversion algorithms has notably improved and more versatile use cases are becoming available with the development of any-to-any (A2A) VC algorithms \cite{freevc,uuvc,casanova2022yourtts,lin2021s2vc}. A2A-VC system denotes a system that can convert speech from any speaker into any unseen speaker.

AutoVC \cite{autoVC} is the first algorithm successfully proposed for A2A-VC. It leverages conditional variational autoencoder (CVAE) to implement an A2A-VC system by learning a one-shot conversion process during training. First, source speech is given to CVAE encoder and pre-trained speaker encoder. Then, the decoder tries to reconstruct the speech with bottleneck feature from the encoder and speaker embedding. To restrict the information from the source, the size of the bottleneck is reduced. Ideally, the remaining bottleneck feature is assumed to have speaker independent linguistic information. The decoder learns how to generate source speaker like speech by combining speaker independent bottleneck features and speaker embedding. In test phase, the conversion process is the same for unseen speakers, except that the target speech is provided to the speaker encoder. Thus, the model can perform one-shot voice conversion if a sufficient number of speakers is seen during the training.

Recently, diffusion models have shown much improved performance in image generation and speech generation \cite{gradtts}\cite{diffusion} and have been applied to the A2A-VC task in DiffVC \cite{diffvc}. Similar to AutoVC, DiffVC consists of an average mel-spectrogram encoder for content embedding, a speaker encoder, and a diffusion-based decoder. The diffusion-based decoder improved converted speech quality by replacing single step generation with multi-step generation. However, since it relies on a single speaker embedding for encoding target speaker characteristics, it often shows limited capability for transferring prosody details such as intonations, loudness and speaking rate (SR), which are crucial factors for the human perception of speaker identity \cite{xu2021influence}. Moreover, it often generates converted speech with unstable fundamental frequency (F0) trajectories when we generated samples.

In this paper, we introduce a highly controllable A2A-VC system with frame-level prosody parameters. To our knowledge, it is the first such model that controls converted speech's F0, loudness, and speaking rate in an A2A-VC system. Prosody transfer is performed in a two-stage process. First, a global prosody conversion step for F0 and SR converts source prosody features into a target like prosody feature. In a second step, user-defined prosody modulation further manipulates prosody before it is merged with speaker and contents embeddings in a prosody conditioning module and then fed to a diffusion-based decoder. We observed that diffusion-based decoder can generate voice with arbitrary speaker identity and prosody. From the experiments, it is shown that the proposed model has an improved mean opinion score (MOS) and better intelligibility compared to the baseline DiffVC model. The proposed model maintains reasonable quality when an appropriate range of F0 or speaking rate modulation is applied.

\section{Background}
\label{sec:format}

A diffusion-based speech generative model was introduced in Grad-TTS\cite{gradtts}. In Grad-TTS, a text encoder first converts input text sequences into a text embedding prior which is an average spectrogram of corresponding text or phoneme. Then, the duration predictor and aligner expands the text embedding prior into realistic length. A diffusion decoder tries to model a diffusion process between the observed distribution and a Gaussian distribution of the text embedding prior. The DiffVC \cite{diffvc} model achieving one-shot VC with high quality leverages both Grad-TTS and AutoVC approaches. It uses an average mel-spectrogram as a content embedding, a speaker embedding from the speaker encoder as well as a diffusion-based decoder model. 

A speaker encoder is used to extract a global speaker embedding for target speech. It is usually implemented with multiple layers which takes mel-spectrogram as an input and predicts speaker labels \cite{dvector}. As a result, the bottleneck feature of the model contains global speaker information. A pre-trained speaker verification network \cite{jia18} is used for the DiffVC model.  
To solely capture linguistic information from speech, an average mel-spectrogram content encoder predicts phonetically divided average mel-spectrogram from the raw spectrogram. The average mel encoder has the same structure as used in DiffVC \cite{diffvc}. It consists of multiple layers of transformer layers.  Like DiffVC training, the encoder is trained separately using phoneme level transcription obtained from the Montreal forced aligner \cite{mfa}. Using alignment information, mel-spectrogram is aggregated for each phoneme duration, and it is used as a target for average mel encoder training. 

Diffusion models are deep generative models that consist of forward and reverse processes\cite{diffusion}. The forward process adds noise to the data while the reverse process tries to recover the original data from the noisy data. In addition, the model learns to remove the noise in multiple steps. As opposed to an ordinary diffusion model, terminal distributions of the DiffVC model are phoneme-dependent single Gaussian distribution. Initially, it starts from the noise added average mel-spectrogram. The diffusion decoder receives speaker embedding from the speaker encoder and noise in the diffused average mel-spectrogram is gradually removed through the generation steps.

\begin{figure*}
    \centering
    \includegraphics[width=0.62\textwidth]{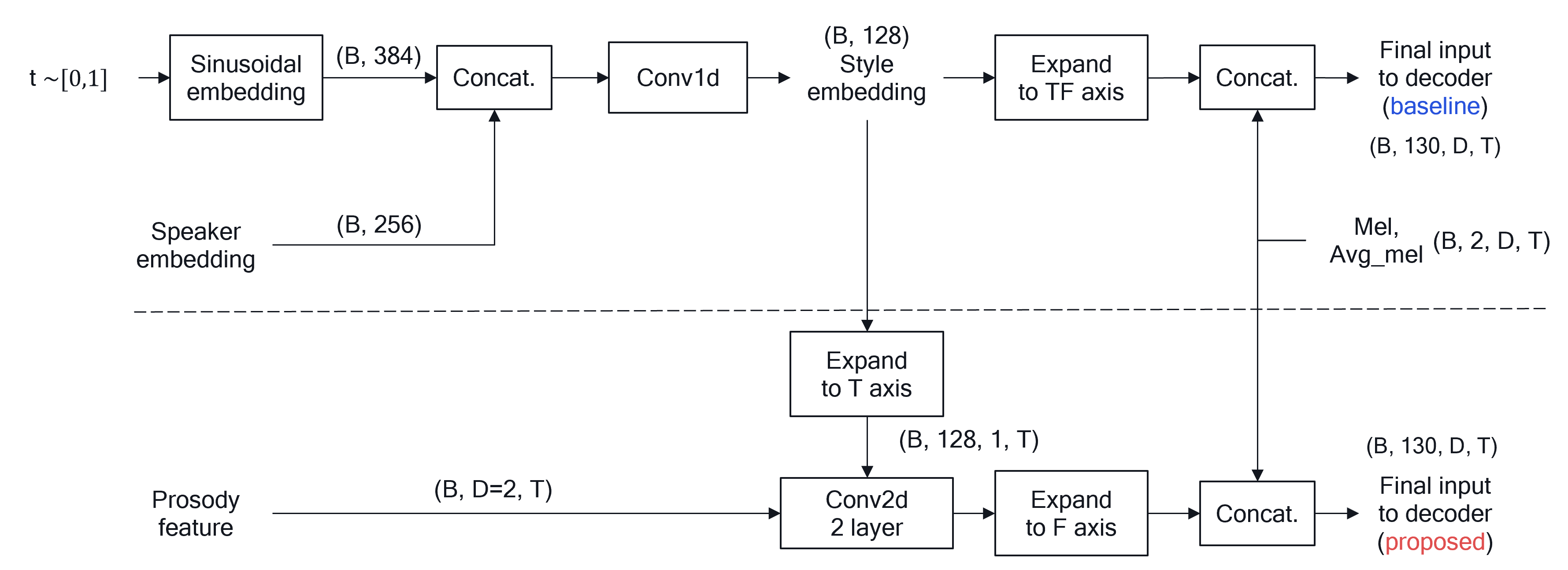}
    \caption{Comparison in conditioning vector between baseline model and proposed prosody conditioning module.}
    \label{fig:condition}
\end{figure*}
\vspace{-10pt}

\section{Proposed model}
\label{sec:pagestyle}

Figure 1 provides an overview of the proposed scheme for training and inference stages. We use a base structure as proposed in \cite{diffvc} including an average mel content encoder, speaker encoder \cite{jia18} and a diffusion decoder. Newly proposed components which are used to integrate the prosody information are highlighted as colored blocks. A prosody extractor extracts frame-level prosody features both in training and inference phase. These features provide additional conditioning for the diffusion decoder. The prosody conversion mechanism consists of 4 modules: Global prosody converter, Prosody modulator, Prosody conditioning and Speaking rate control. With frame-level F0 and energy, the proposed model can generate any speaker's voice with more controllability thereby allowing intonation variation and speaking speed modification. The following sections provide a detailed description of how we integrated prosody features. 

\subsection{Prosody extractor}
The prosody extractor extracts logF0, log energy, and speaking rate to capture prosody features from the source and target speech. For the F0 extraction, we first apply a high-pass filter: a Butterworth filter with a 50 Hz cut-off frequency is used to prevent F0 tracking errors. Then, the F0 trajectory is obtained using the librosa\cite{librosa} toolkit, and F0 value for unvoiced region is set to zero. For the energy, the squared sum of the samples in each frame is measured, and its logarithm is taken. For the speaking rate information, we use Hubert model \cite{hubert} to get the [unit, duration] pairs from speech, the average speaking rate is obtained as inverse of average duration length. In this way, the speaking rate can be measured without automatic speech recognition (ASR). 

\subsection{Training of diffusion decoder with frame-level prosody conditioning}
During training, as shown in  Figure 1. (a), randomly paired source speeches are separately given to the speaker encoder and remaining encoders. The prosody embedding is merged with other embeddings in the prosody conditioning module. Then, the U-Net based diffusion decoder generates mel-spectrogram given noise scheduler \textit{t} and all encoder embeddings. The decoder is trained to minimize the MSE loss between ground truth noise and estimated noise from the generated mel-spectrogram.

Fig. 2 demonstrates the structure of the proposed prosody conditioning module and original conditioning method in DiffVC. In the DiffVC, step information \textit{t} and speaker embedding were concatenated and style embedding is obtained after convolution layer. Then, its time and frequency axis were expanded to have the same dimensionality as mel-spectrogram. The prosody conditioning module takes style embedding and frame-level prosody features as inputs and integrates them into a time-varying condition feature. The merging layer consists of a two-layer convolutional network accepting logF0 and log energy as prosody features and style embeddings that are expanded in time axis. Then, followed by a frequency axis expansion, they are concatenated to mel-spectrograms. The concatenated vector is used as an input to the decoder.

\subsection{Inference scheme with prosody modulation }
During inference, as shown in the Figure 1. (b), target speaker prosody information is sent to global prosody converter and prosody modulator. The global prosody converter modifies source prosody feature to follow target speech prosody. The global prosody converter performs mean F0 transfer and computes speaking conversion rate. F0 mean transfer modifies source F0 to have target’s mean F0 as represented in Equation 1. 
$F0_{new}=F0_{src} + (\mu_{trg} - \mu_{src}) * V$, where V is voiced flag. 
For the speaking conversion rate, $R_c$, we used Hubert model to compute the average duration of speech tokens of source, $D_{src}$ and target $D_{trg}$ speech. The inverse ratio between $D_{src}$ and $D_{trg}$ is computed as follows. 
$R_c = \frac{D_{src}}{D_{trg}}$.

In the prosody modulator, a user may add changes after prosody conversion. The user defined prosody modulator is either global modulation or frame-level modulation. In the case of global modulation, for example, it is possible to increase 3 semitones for all frames. In case of frame-level modulation, user can make arbitrary modification to the F0, such as delicate intonation control is possible.

The modulated prosody features are combined with target speaker embedding and source contents embedding in the prosody conditioning module. Then, the diffused average mel-spectrogram is gradually converted to clean target speech like mel-spectrogram in diffusion decoder with the embeddings from the prosody conditioning module. 
Optionally, speaking rate can be controlled after the mel-spectrogram generation via re-sampling method. In the speaking rate control module, the $R_c$ obtained from the global prosody converter is used as the re-sampling ratio for the mel-spectrogram output from the diffusion decoder. The range of the conversion rate is limited to [0.66, 1.33] to prevent severe degradation. Finally, Hifi-GAN neural vocoder is used to obtain the output waveform.

\vspace{-10pt}
\section{Experiments}
\label{sec:typestyle}
\vspace{-10pt}

\begin{table}[t]
\caption{Experiment result}
\centering
\begin{tabular}{ccccc}
\toprule
                  & \multicolumn{4}{c}{\textbf{Metric}}               \\
\midrule
\textbf{System}         & \textbf{MOS} & \textbf{NISQA}  & \textbf{SECS} & \textbf{CER} \\
\textbf{DiffVC}         & 4.01$\pm$0.15  &  \textbf{3.44}   & \textbf{0.324}   & 5.26             \\
\textbf{Proposed}       & \textbf{4.31$\pm$0.13}            &  3.37            & 0.321         & \textbf{4.18}             \\
\textbf{Proposed(SR)}    & 4.15$\pm$0.14          &   3.28             & 0.311         & 4.50             \\
\bottomrule
\end{tabular}
\end{table}

\begin{table}[t]
\caption{VC with F0, and speaking rate modulation result}
\centering
\begin{tabular}{cccc}
\toprule
                  & \multicolumn{3}{c}{\textbf{Metric}}               \\
\midrule
\textbf{F0 modulation}   & \textbf{NISQA} & \textbf{SECS} & \textbf{CER(\%)}    \\
\midrule
\textbf{-0.50}          & 3.37           & 0.274         & 4.90                 \\
\textbf{-0.25}          & 3.36           & 0.299         & 4.41                 \\
\textbf{Global F0 mod.} & 3.37           & \textbf{0.321}         & 4.18                 \\
\textbf{0.25}           & 3.45           & 0.312         & \textbf{4.14}                 \\
\textbf{0.50}           & \textbf{3.57}           & 0.294         & 4.61                 \\
\midrule
\textbf{SR modulation}      & \textbf{NISQA} & \textbf{SECS} & \textbf{CER(\%)} \\
\midrule
\textbf{x0.66}              & 3.18           & 0.259         & 6.70             \\
\textbf{x0.75}              & \textbf{3.31}           & 0.284         & 4.89             \\
\textbf{Global SR conv.}        & 3.28           & \textbf{0.311}         & 4.50             \\
\textbf{x1.20}              & 3.08           & 0.281         & \textbf{4.44}             \\
\textbf{x1.33}              & 2.95           & 0.266         & 5.02             \\
\bottomrule
\end{tabular}
\end{table}

\subsection{Experiment settings}
We used the LibriTTS\cite{Libritts} database for the training and evaluation. Randomly selected 10 speakers from the LibriTTS \textit{train-clean-360h} set were used for evaluation. The \textit{train-clean-100h}, and remaining \textit{train-clean-360h} sets were used as a training set.  Adam optimizer \cite{adam} with an 1e-4 learning rate was used for the encoder and decoder training. The averaged mel-spectrogram encoder was trained with phoneme transcription from the Montreal forced aligner \cite{mfa}. The diffusion-based decoder was trained for 100 epochs; speaker encoder and content encoder were frozen during this phase. The condition merging layer and diffusion model were fine-tuned for 5 epochs. For the evaluation, there are 90 conversion pairs between 10 speakers from train-clean-360h. For each speaker conversion pair, 30 pairs of [source, target] were generated and used for all systems, which results in 2,700 pairs in total for single system evaluation. For the generation, 30 steps are used as in the DiffVC model.

Three objective measures; NISQA (Non-Intrusive Speech Quality Assessment) \cite{nisqa} score, speaker embedding cosine similarity (SECS), and the character error rate (CER) were used. NISQA model predicts a 5-scale MOS for overall synthesis quality. SECS is used for measuring the speaker similarity between target and converted voice. We used ECAPA-TDNN \cite{ecapa} as the speaker embedding extractor. The Wav2Vec2.0-based ASR model \cite{wav2vec2} was used to measure the CER between source and converted voice for intelligibility of the converted speech. For the perceptual evaluation, we conducted MUSHRA \cite{mushra} test to obtain the mean opinion score (MOS) \cite{MOS}.The 7 human listeners participated to evaluate the overall quality of 25 randomly selected samples.

\subsection{Experimental results}
Table 1 summarizes the performance of the baseline system and the proposed systems. Overall, the proposed model without speaking rate (SR) conversion has comparable naturalness and slightly better intelligibility compared to the baseline DiffVC model. The proposed system produces more intelligible speech with explicit prosody information with the lowest 4.18\% CER among the systems. For SECS, both systems have similar performance 0.324, and 0.321 for baseline and the proposed model respectively. 

With regard to naturalness as measured subjectively by a human panel, the proposed model showed the best MOS which is a statistically significant higher score than the DiffVC model result. However, in terms of the objective metric NISQA, the baseline model is roughly in line with the proposed model. Considering both objective and subjective measures, the two systems have comparable quality. Regarding the SR module effect, all evaluation metrics are degraded compared to without SR conversion case. It enables model speaking rate controllability but results in no performance gain. It seems that SR conversion at a lower level such as phoneme or word units is needed rather than the global level to maintain the naturalness of the system. 

The effect of F0 modulation and SR conversion proposed in the paper is analyzed in Table 2. F0 is modulated from -0.5 octave to 0.5 octave with 0.25 octave interval. According to the experimental results, it was confirmed that the performance of SECS and CER deteriorated as more modulation was applied. CER is not affected significantly, but SECS showed a larger degradation, indicating that the change in F0 is related to the speaker identity. However, considering that the threshold of the ECAPA-TDNN model used in the experiment is set to 0.25, it can be seen that on average, the speaker identity of F0 modulated speech still sounds like a target speaker. The impact of global SR conversion is analyzed by changing the SR from 0.66 to 1.33. Similar to F0 modulation result, a decrease in quality was observed as the change increased. However, the drop was larger than that of F0 modulation.

\vspace{-10pt}
\section{Conclusion}
\label{sec:majhead}
A highly controllable voice conversion system is proposed that explicitly feeds prosody features into a diffusion-based model. With the proposed system, comparable speech quality and improved intelligibility was achieved compared to the baseline model. Moreover, the approach enables a considerable range of prosody modulation with small degradation of speech quality. The proposed prosody control feature can be used not only for voice conversion but also for speech manipulation, such as altering intonations. In the future, we will extend our work to a prosody conversion model which converts a raw prosody feature into new trajectories with specified styles or speaker identities.
\vfill\pagebreak
\vfill\pagebreak

\bibliographystyle{IEEEbib}
\bibliography{main}

\end{document}